\documentclass[english,showpacs]{revtex4}
\usepackage[T1]{fontenc}
\usepackage[latin9]{inputenc}
\usepackage{amsthm}
\usepackage{amsmath}
\usepackage{graphicx}
\usepackage{esint}

\usepackage{babel}

\begin{document}

\title{Light hadron production in $B_{c}\to J/\psi+X$ decays}

\author{A. K. Likhoded}

\email{Anatolii.Likhoded@ihep.ru}

\affiliation{Institute for High Energy Physics, Protvino, Russia}

\author{A. V. Luchinsky}

\email{Alexey.Luchinsky@ihep.ru}

\affiliation{Institute for High Energy Physics, Protvino, Russia}

\begin{abstract}
Decays of ground state $B_{c}$-meson $B_{c}\to J/\psi+n\pi$ are
considered. Using existing parametrizations for $B_{c}\to J/\psi$
form-factors and $W^{*}\to n\pi$ spectral functions we calculate branching
fractions and transferred momentum distributions of $B_{c}\to J/\psi+n\pi$
decays for $n=1,2,3,4$. Inclusive decays $B_{c}\to J/\psi+\bar{u}d$
and polarization asymmetries of final charmonium are also investigated.
Presented in our article results can be used to study form-factors
of $B_{c}\to J/\psi$ transitions, $\pi$-meson system spectral functions
and give the opportunity to check the factorization theorem. 
\end{abstract}
\maketitle

\section{Introduction}

Recent measurements of $B_{c}$-meson mass and lifetime in CDF \cite{Aaltonen2007gv}
and D0 \cite{Abazov2008rb} experiments allow us to hope that more
detailed investigation of this particle on LHC collider, where about
$10^{10}$ $B_{c}$-events per year are expected, would clarify mechanisms
of $B_{c}$ production and decay modes. Currently only products of
$B_{c}$-meson production cross section and branching fractions of
decays $B_{c}\to J/\psi\pi$, $J/\psi\ell\nu$ are known experimentally.
For example, the following ratios are measured \cite{Papadimitriou2006}:\begin{eqnarray*}
\frac{\sigma_{B_{c}}Br\left(B_{c}\to J/\psi e^{+}\nu_{e}\right)}{\sigma_{B}Br\left(B_{c}\to J/\psi K\right)} & = & 0.282\pm0.038\pm0.074\end{eqnarray*}
for positron in the final state and\begin{eqnarray*}
\frac{\sigma_{B_{c}}Br\left(B_{c}\to J/\psi\mu^{+}\nu_{\mu}\right)}{\sigma_{B}Br\left(B_{c}\to J/\psi K\right)} & = & 0.249\pm0.045_{-0.076}^{+0.107}\end{eqnarray*}
for muon. These ratios are about an order of magnitude higher than
the theoretical predictions based on current estimates of $B_{c}$-meson
production cross section and branching fraction $Br(B_{c}\to J/\psi\ell\nu)\approx2\%$
\cite{Gershtein1995}. The mode $B_{c}\to J/\psi\pi$ was used mainly
to determine precisely $B_{c}$-meson mass. No information on production
cross section, decay branching fraction, and even the product of these
quantities was determined in this experiment.

Investigation of other $B_{c}$-meson decay channels and determination
of their branching fractions will be one of interesting tasks of future
experiments on LHC. Weak $B_{c}$ decays can be caused by decays of
both constituent quarks. Dominant are $c$-quark decay modes, which
amount to $\sim70\%$ of all $B_{c}$-meson decays. Unfortunately,
none of such reactions were observed, although large branching fractions
are expected for some of these decay modes (for example, for $B_{c}\to B_{s}\rho$
we have approximately 16\% branching fraction). Mentioned above decays
$B_{c}\to J/\psi\ell\nu$ and $B_{c}\to J/\psi\pi$ are examples of
other class, caused by $b$-quark decay. Total branching fraction
of this process is about 20\%.

In the present paper we will fill the gap in existing theoretical
predictions of $B_{c}$-meson decay branching fractions \cite{Gershtein1995,Gershtein1997qy,Kiselev2000nf,Kiselev2000pp}
and consider multi-particle processes $B_{c}\to J/\psi+n\pi$ with
$n=1,2,3,4$. These reactions are caused by weak $b$-quark decay
$b\to cW^{*}\to c\bar{u}d$ and clean analogy with similar $\tau$-lepton
decays ($\tau\to\nu_{\tau}+n\pi$) can be easily seen. This analogy
allows us to use existing experimental data on $\tau$-lepton decays
and give reliable predictions of $B_{c}\to J/\psi+n\pi$ branching
fractions.

In the next section we give analytical expressions for distributions
of $B_{c}\to J/\psi+n\pi$ decays branching fractions over invariant
mass of the light hadron system and study different asymmetries of
final $J/\psi$-meson polarization as a function of this kinematic
variable. In section III we use existing experimental data on $\tau$-lepton
decays calculate branching fractions of $B_{c}\to J/\psi+n\pi$ decays
for $n=1,2,3,4$. In section IV inclusive reaction $B_{c}\to J/\psi\bar{u}d$
is considered in connection with duality relation. Short results of
our work are given in the final section.

\section{Analytic Results}

\begin{figure}[b]
\begin{centering}
\includegraphics{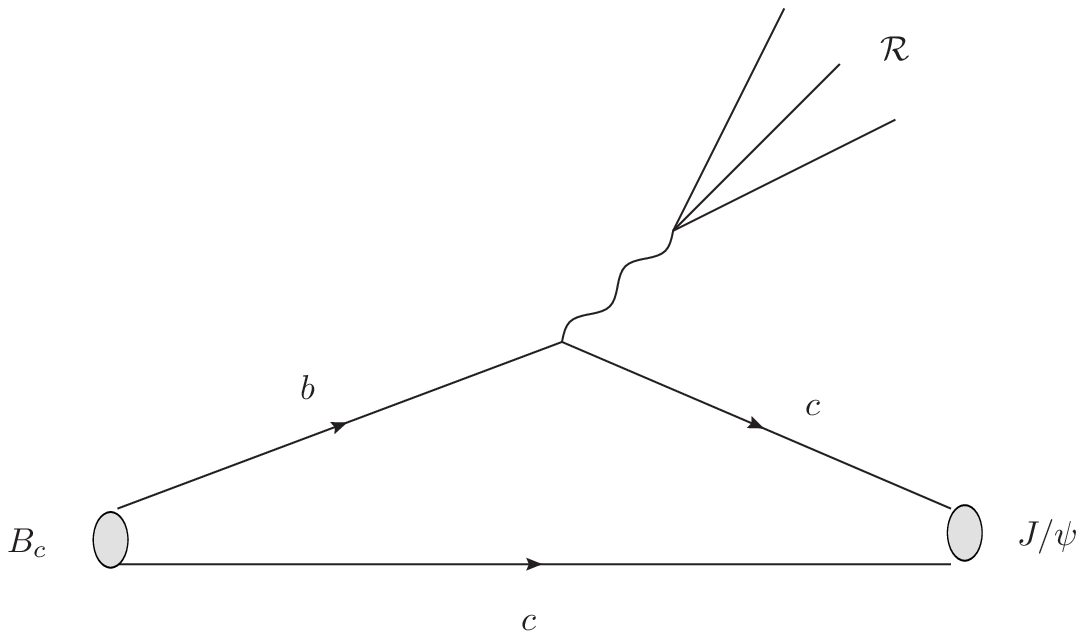} 
\par\end{centering}

\caption{$B_{c}\to J/\psi+\mathcal{R}$\label{fig:diag}}

\end{figure}

$B_c$-meson decays into light hadrons with vector charmonium $J/\psi$ production are caused by $b$-quark decay $b\to W^{*} \to c \bar{u} d$ (see diagram shown in fig.\ref{fig:diag}). The effective lagrangian of the latter process reads
\begin{eqnarray*}
 \mathcal{H}_\mathrm{eff} &=& 
	\frac{G_F}{2\sqrt{2}}V_{cb}V_{ud}^*\left[
		C_{+}(\mu) O_{+} + C_{-}(\mu) O_{-}
	\right],
\end{eqnarray*}
where $G_{F}$ is Fermi coupling constant, $V_{ij}$ are the elements of CKM mixing matrix, $C_{\pm}(\mu)$ are Wilson coefficients, that take into account higher QCD corrections and operators $O_{\pm}$ are defined according to
\begin{eqnarray*}
 O_{\pm} &=&
	(\bar d_i u_j)_{V-A}(\bar c_i b_j)_{V-A} \pm (\bar d_j u_i)_{V-A}(\bar c_i b_j)_{V-A}.
\end{eqnarray*}
In this expression $i,j$ are color indexes of quarks and $(\bar q_1 q_2)_{V-A}=\bar q_1\gamma_\mu(1-\gamma_5)q_2$. Since in our decays light quark pair should be in color-singlet state, the amplitude of the considered here processes is proportional to
\begin{eqnarray*}
 a_1(\mu) &=&
	\frac{1}{2N_c}\left[ (N_c-1)C_{+}(\mu) + (N_c-1) C_{-}(\mu) \right]
\end{eqnarray*}
If QCD corrections are neglected, one  should set $a_1(\mu)=1$. Leading logarithmic strong corrections lead to dependence of this coefficient on the renormalization scale $\mu$ \cite{Buchalla1996}, and on $\mu\sim m_{b}$ it is equal to
\begin{eqnarray*}
a_1\left(m_{b}\right) & = & 1.17.
\end{eqnarray*}

The matrix element of the decay $B_{c}\to J/\psi+\mathcal{R}$, where
$\mathcal{R}$ is some set of light hadrons, has the form
\begin{eqnarray}
\mathcal{M}\left[B_{c}\to W^{*}J/\psi\to\mathcal{R}J/\psi\right] & = & 
	\frac{G_{F}V_{cb}}{\sqrt{2}}a_1\mathcal{H}^{\mu}\epsilon_{\mu}^{\mathcal{R}}.\label{eq:Matr}
\end{eqnarray}
In this expression $\epsilon^{\mathcal{R}}$ is the effective polarization
vector of virtual $W$-boson and

\begin{eqnarray*}
\mathcal{H}_{\mu} & = & \left\langle J/\psi\left|\bar{c}\gamma_{\mu}\left(1-\gamma_{5}\right)b\right|B_{c}\right\rangle =\mathcal{V}_{\mu}-\mathcal{A}_{\mu}.\end{eqnarray*}

Vector and axial currents are equal to

\begin{eqnarray*}
\mathcal{V}_{\mu} & = & \left\langle J/\psi\left|\bar{c}\gamma_{\mu}b\right|B_{c}\right\rangle =i\epsilon^{\mu\nu\alpha\beta}\epsilon_{\nu}^{\psi}\left(p+k\right)_{\alpha}q_{\beta}F_{V}\left(q^{2}\right),\\
\mathcal{A}_{\mu} & = & \left\langle J/\psi\left|\bar{c}\gamma_{\mu}\gamma_{5}b\right|B_{c}\right\rangle =\epsilon_{\mu}^{\psi}F_{0}^{A}\left(q^{2}\right)+\left(\epsilon^{\psi}p\right)\left(p+k\right)_{\mu}F_{+}^{A}\left(q^{2}\right)+\left(\epsilon^{\psi}p\right)q_{\mu}F_{-}^{A}\left(q^{2}\right),\end{eqnarray*}
where $p$ and $k$ are the momenta of $B_{c}$- and $J/\psi$-mesons,
$q=p-k$ is the momentum of virtual $W$-boson, and $F_{V}(q^{2})$,
$F_{0,\pm}^{A}(q^{2})$ are form-factors of $B_{c}\to J/\psi W^{*}$
decay. Due to vector current conservation and partial axial current
conservation the contribution of the form-factor $F_{-}^{A}$ are
suppressed by small factor $\sim(m_{u}+m_{d})^{2}/M_{B_{c}}^{2}$,
so we will neglect it in the following.

One can use different approaches when deriving the form of the form-factors
$F(q^{2})$. First of all, it is clear, that quark velocity in heavy
quarkonia is small in comparison with $c$, so one can describe heavy
quarkonia in the terms of non-relativistic wave-functions. This fact
was used on the so called Quark Models \cite{Kiselev1992tx,Gershtein1995,Gershtein1997qy,Colangelo1999zn,Kiselev1999sc,Ivanov2005fd,Hernandez2006gt,Ivanov2006ni}.
In the following we will refer to this set of form-factors as \emph{QM}.The
speed of the final charmonium in $B_{c}$-meson rest frame, on the
other hand, is large, so one can expand the amplitude of the considered
here process in the powers of small parameter $M_{J/\psi}/M_{B_{c}}$,
as it was done in papers \cite{Anisimov1998xv,Anisimov1998uk,Huang2008zg,Choi2009,Choi2009ai,Wang2009mi}.
In what follows, we will refer to this set of form-factors as \emph{LC}.
One can also use 3-point QCD sum rules to obtain the information on
$B_{c}\to J/\psi W^{*}$form-factors \cite{Gershtein1995,Kiselev1999sc,Kiselev2002vz,Azizi2008vt}
(\emph{SR}).

In our paper we use the following simple parametrization of form-factors

\begin{eqnarray*}
F\left(q^{2}\right) & = & \frac{F\left(0\right)}{1-q^{2}/M_{pole}^{2}},\end{eqnarray*}
where numerical values of parameters $F_{i}(0)$ and $M_{pole}$ are
presented in table \ref{tab:FF}.

\begin{table}
\begin{centering}
\begin{tabular}{|c|c|c|c|c|}
\hline 
\multicolumn{2}{|c|}{} & SR & QM & LC\tabularnewline
\hline
\hline 
$F_{V}$ & $F_{V}(0)$, $\text{GeV}^{-1}$ & 0.11 & 0.10 & 0.08\tabularnewline
\cline{2-5} 
 & $M_{pole}$, GeV & 4.5 & 4.5 & 4.5\tabularnewline
\hline
\hline 
$F_{0}^{A}$ & $F_{0}^{A}(0)$, GeV & 5.9 & 6.2 & 4.7\tabularnewline
\cline{2-5} 
 & $M_{pole}$, GeV & 4.5 & 4.5 & 6.4\tabularnewline
\hline
\hline 
$F_{+}^{A}$ & $F_{+}^{A}(0)$, $\text{GeV}^{-1}$ & -0.074 & -0.70 & -0.047\tabularnewline
\cline{2-5} 
 & $M_{pole}$, GeV & 4.5 & 4.5 & 5.9\tabularnewline
\hline
\end{tabular}
\par\end{centering}

\caption{Parameters of $B_{c}$-meson form-factors\label{tab:FF}}

\end{table}

The width of the $B_{c}\to J/\psi\mathcal{R}$ decay is

\begin{eqnarray*}
d\Gamma\left(B_{c}\to J/\psi\mathcal{R}\right) & = & \frac{1}{2M}\frac{G_{F}^{2}V_{cb}^{2}}{2}a_{1}^{2}\mathcal{H}^{\mu}\mathcal{H}^{*\nu}\epsilon_{\mu}\epsilon_{\nu}^{*\mathcal{R}}d\Phi\left(B_{c}\to J/\psi\mathcal{R}\right),\end{eqnarray*}
where Lorentz-invariant phase space is defined according to

\begin{eqnarray*}
d\Phi\left(Q\to p_{1}\dots p_{n}\right) & = & (2\pi)^{4}\delta^{4}\left(Q-\sum p_{i}\right)\prod\frac{d^{3}p_{i}}{2E_{i}(2\pi)^{3}}.\end{eqnarray*}
It is well known, that the following recurrent expression holds for
this phase space:

\begin{eqnarray*}
d\Phi\left(B_{c}\to J/\psi\mathcal{R}\right) & = & \frac{dq^{2}}{2\pi}d\Phi\left(B_{c}\to J/\psi W^{*}\right)d\Phi\left(W^{*}\to\mathcal{R}\right).\end{eqnarray*}
Using this expression one can perform the integration over phase space
of the final state $\mathcal{R}$:

\begin{eqnarray*}
\frac{1}{2\pi}\int d\Phi\left(W^{*}\to\mathcal{R}\right)\epsilon_{\mu}^{\mathcal{R}}\epsilon_{\nu}^{\mathcal{R}*} & = & \left(q_{\mu}q_{\nu}-q^{2}g_{\mu\nu}\right)\rho_{T}^{\mathcal{R}}\left(q^{2}\right)+q_{\mu}q_{\nu}\rho_{L}^{\mathcal{R}}\left(q^{2}\right),\end{eqnarray*}
where spectral functions $\rho_{T,L}^{\mathcal{R}}\left(q^{2}\right)$
are universal and can be determined from theoretical and experimental
analysis of some other processes, for example $\tau\to\nu_{\tau}\mathcal{R}$
decay or electron-positron annihilation $e^{+}e^{-}\to\mathcal{R}$.
Due to vector current conservation and partial axial current conservation
spectral function $\rho_{L}^{\mathcal{R}}$ is negligible on almost
whole kinematical region, so we will neglect is in our paper. Explicit
expressions for spectral function $\rho_{T}^{\mathcal{R}}$ for different
final states $\mathcal{R}$ are given in the next section.

Differential distributions of longitudinally and transversely polarized
$J/\psi$-meson in $B_{c}\to J/\psi+\mathcal{R}$ decays can easily
be obtained from presented above expressions. In the case of longitudinal
polarization the polarization vector $\epsilon^{\psi}$ is equal to\begin{eqnarray*}
\epsilon_{\mu}^{\psi}\left(\lambda=0\right) & = & \frac{M}{2M_{V}}\left\{ \beta,0,0,\frac{M^{2}+M_{V}^{2}-q^{2}}{M^{2}}\right\} ,\end{eqnarray*}
where $z$-axes is chosen in the direction of $J/\psi$ movement,
$M$ and $M_{v}$ are $B_{c}$- and $J/\psi$-meson masses and\begin{eqnarray*}
\beta & = & \sqrt{\frac{\left(M+M_{V}\right)^{2}-q^{2}}{M^{2}}}\sqrt{\frac{\left(M-M_{V}\right)^{2}-q^{2}}{M^{2}}}.\end{eqnarray*}
Differential distribution has the form\begin{eqnarray*}
\frac{d\Gamma\left[B_{c}\to J/\psi_{\lambda=0}+\mathcal{R}\right]}{dq^{2}} & = & \frac{G_{F}^{2}M^{3}V_{cb}^{2}a_{1}^{2}\beta}{128\pi M_{V}^{2}}\rho_{T}^{\mathcal{R}}\left(q^{2}\right)\frac{M^{4}}{4M_{V}^{2}}\left\{ \left(\beta^{2}+\frac{4M_{V}^{2}q^{2}}{M^{4}}\right)\left|F_{0}^{A}\right|^{2}+M^{4}\beta^{4}\left|F_{+}^{A}\right|^{2}\right.\\
 & + & \left.2\beta^{2}\left(M^{2}-M_{V}^{2}-q^{2}\right)F_{0}^{A}F_{+}^{A}\right\} .\end{eqnarray*}
In the case of transversely polarized vector meson $\epsilon_{\mu}^{\psi}$
has the form\begin{eqnarray*}
\epsilon_{\mu}^{\psi}\left(\lambda=\pm1\right) & = & \left\{ 0,\frac{1}{\sqrt{2}},\frac{\pm i}{\sqrt{2}},0\right\} ,\end{eqnarray*}
and the corresponding differential distribution is\begin{eqnarray*}
\frac{d\Gamma\left[B_{c}\to J/\psi_{\lambda=\pm1}+\mathcal{R}\right]}{dq^{2}} & = & \frac{G_{F}^{2}V_{cb}^{2}}{32\pi M}a_{1}^{2}\beta q^{2}\rho_{T}^{\mathcal{R}}\left(q^{2}\right)\left\{ \left|F_{0}^{A}\right|^{2}+M^{4}\beta^{2}\left|F_{V}\right|^{2}\pm\frac{2\beta M^{2}}{M_{V}^{2}}\mbox{Re}\left(F_{0}^{A}F_{V}\right)\right\} .\end{eqnarray*}
It should be stressed, that the above expressions are universal and
spectral function $\rho_{T}^{\mathcal{R}}\left(s\right)$ depends
on the final state $\mathcal{\mathcal{R}}$.

\begin{figure}
\begin{centering}
\includegraphics[scale=0.7]{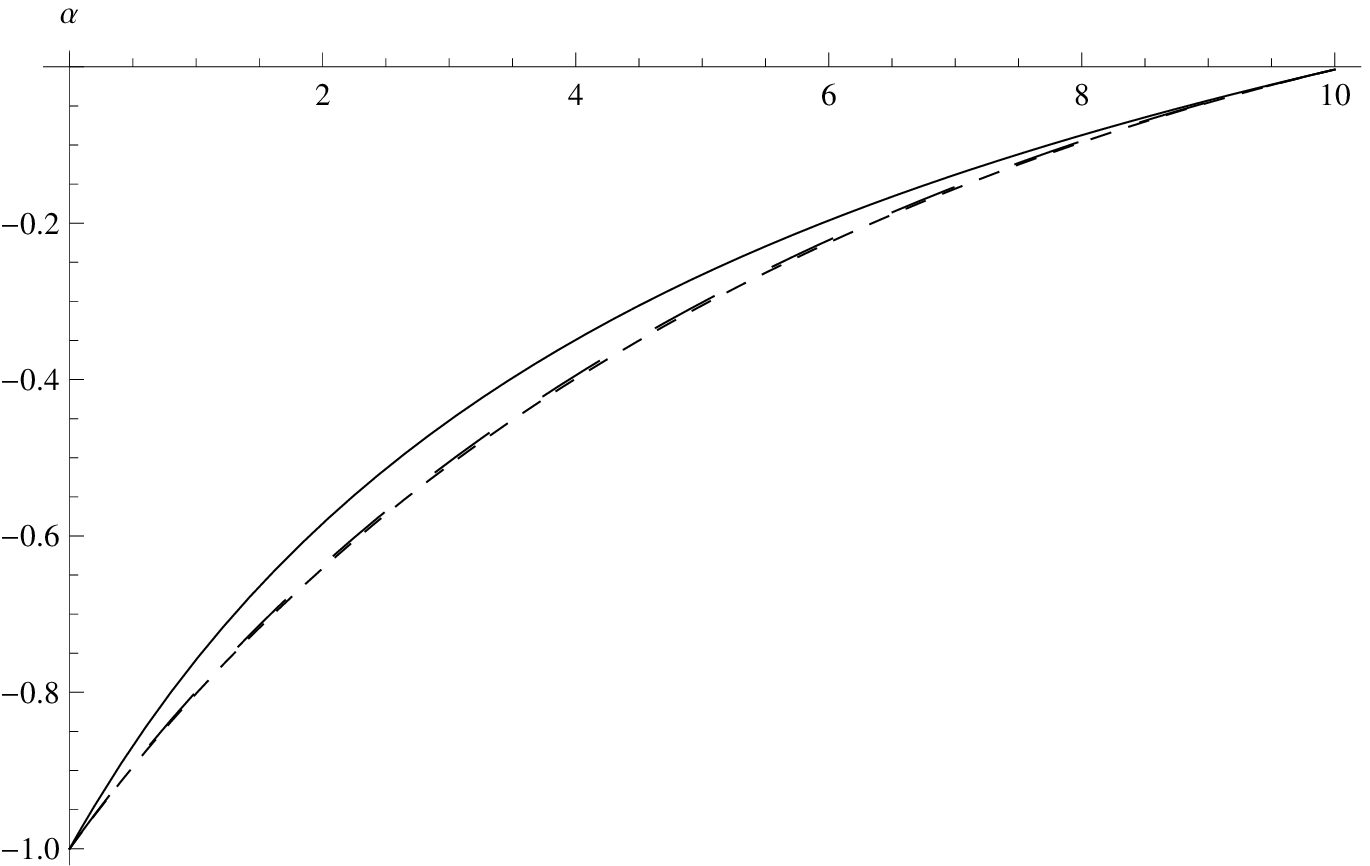}
\par\end{centering}

\caption{Polarization asymmetry $\alpha$ of final $J/\psi$-meson in $B_{c}\to J/\psi+\mathcal{R}$
decays as a function of squared transferred momentum $q^{2}$ (in
$\mbox{GeV}^{2}$). Solid, dashed and dot-dashed lines stand for SR
\cite{Kiselev2000pp,Kiselev1999sc}, QM \cite{Gershtein1995}, and
LC \cite{Wang2009mi} respectively \label{fig:alpha}}

\end{figure}

\begin{figure}
\begin{centering}
\includegraphics{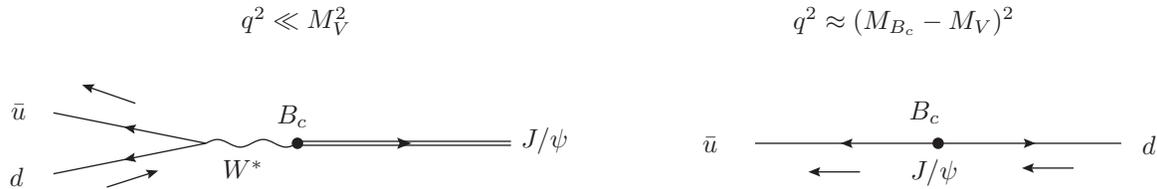}
\par\end{centering}

\caption{Kinematics of $B_{c}\to J/\psi u\bar{d}$ decay\label{fig:helicities}}

\end{figure}

If the polarization if final vector meson is not observed, the $q^{2}$-distribution
is, obviously,\begin{eqnarray}
\frac{d\Gamma\left[B_{c}\to J/\psi+\mathcal{R}\right]}{dq^{2}} & = & \sum_{\lambda=0,\pm1}\frac{d\Gamma\left[B_{c}\to J/\psi_{\lambda}+\mathcal{R}\right]}{dq^{2}}.\label{eq:dGamma}\end{eqnarray}
It is also useful to study some polarization asymmetries. For example,
polarization degree $\alpha$ is defined according to\begin{eqnarray*}
\alpha & = & \frac{d\Gamma_{\lambda=+1}+d\Gamma_{\lambda=-1}-2d\Gamma_{\lambda=0}}{d\Gamma_{\lambda=+1}+d\Gamma_{\lambda=-1}+2d\Gamma_{\lambda=0}}.\end{eqnarray*}
Production of transversely polarized, longitudinally polarized and
unpolarized $J/\psi$-meson corresponds to $\alpha=1$, $\alpha=-1$
and $\alpha=0$ respectively. We would like to note, that in the framework
of factorization model this asymmetry does not depend on final state
$\mathcal{R}$. So, experimental investigation of this asymmetry can
be used for determination of $B_{c}$-meson form factors and test
of QCD factorization. In fig.\ref{fig:alpha} we show $q^{2}$-dependence
of this asymmetry for different sets of $B_{c}$-meson form-factors.
One can easily explain qualitatively the behavior of these curves.
Let us consider $q^{2}$-dependence of asymmetry $\alpha$ in $B_{c}\to J/\psi\bar{u}d$
decay. At low $q^{2}$ the direction of $\bar{u}$- and $d$-quarks
momenta in $B_{c}$-meson rest frame will be close to each other and
opposite to the direction of the momentum of $J/\psi$-meson. The
spin of light $\bar{u}$-antiquark ($d$-quark) is directed along
(opposite to) its momentum (see fig.\ref{fig:helicities}a), so quark-antiquark
pair has $\lambda=0$ projection on $Oz$ axis. From angular momentum
conservation it follows, that $J/\psi$-meson should also be longitudinally
polarized. This can be observed in figure \ref{fig:alpha}, where
at low $q^{2}$ we have $\alpha=-1$ for all sets of $B_{c}$-meson
form-factors. In high $q^{2}$-region, on the contrary, direction
of quark and antiquark momenta are opposite to each other and $J/\psi$-meson
stay at rest in $B_{c}$-meson rest frame (see fig.\ref{fig:helicities}b).
As a result, final $J/\psi$-meson is unpolarized in this region and
$\alpha=0$.

\begin{figure}[b]
\begin{centering}
\includegraphics[scale=0.7]{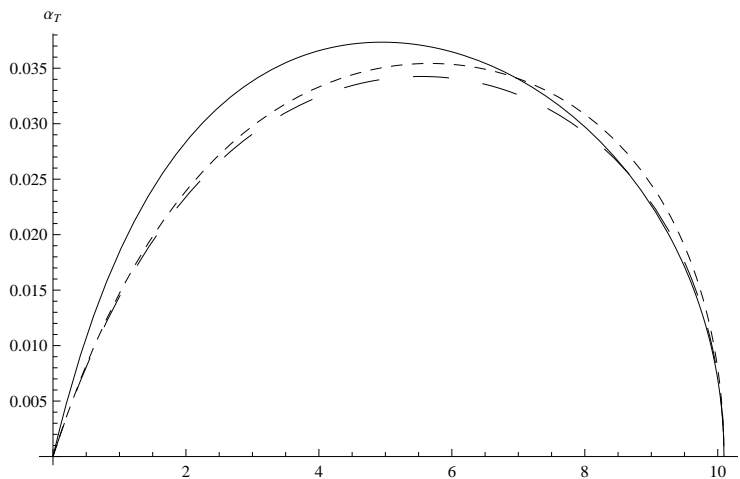}
\par\end{centering}

\caption{Transverse polarization asymmetry $\alpha_{T}$ of final $J/\psi$-meson
in $B_{c}\to J/\psi+\mathcal{R}$ decays as a function of squared
transferred momentum $q^{2}$ (in $\mbox{GeV}^{2}$). Solid, dashed
and dot-dashed lines stand for SR \cite{Kiselev2000pp,Kiselev1999sc},
QM \cite{Gershtein1995}, and LC \cite{Wang2009mi} respectively \label{fig:alphaT}}

\end{figure}

Another example is transverse asymmetry\begin{eqnarray*}
\alpha_{T} & = & \frac{d\Gamma_{\lambda=1}-d\Gamma_{\lambda=-1}}{d\Gamma}.\end{eqnarray*}
This asymmetry also depends only on $B_{c}$-meson form-factors and
its dependence on squared transferred momentum is shown in fig.\ref{fig:alphaT}.

\section{Exclusive Decays}

In this section we present differential widths and branching fractions
of the decays $B_{c}\to J/\psi+n\pi$ using presented above universal
formula (\ref{eq:dGamma}) and specific expressions for spectral function
$\rho_{T}^{\mathcal{R}}\left(q^{2}\right)$.

\subsection{$B_{c}\to J/\psi\pi$}

Let us first of all consider two-particle decays $B_{c}\to J/\psi\pi$
and $B_{c}\to J/\psi\rho$.

In the case of $B_{c}\to J/\psi\pi$ decay the $W^{*}\to\pi$ transition
is expressed through leptonic constant $f_{\pi}$:

\begin{eqnarray}
\left\langle \pi\left|\bar{u}\gamma_{\mu}\gamma_{5}d\right|0\right\rangle  & = & \sqrt{2}f_{\pi}q_{\mu}.\label{eq:pi}\end{eqnarray}
The numerical value of this constant can be determined from $\pi\to\mu\nu_{\mu}$
decay width: $f_{\pi}\approx140$ MeV. The spectral function, that
corresponds to vertex (\ref{eq:pi}) is

\begin{eqnarray*}
\rho_{T}^{\pi}\left(q^{2}\right) & = & 2f_{\pi}^{2}\delta\left(q^{2}\right).\end{eqnarray*}
Using this spectral function it is easy to obtain the following values
of $B_{c}\to J/\psi\pi$ decay branching fractions for different sets
of form-factors:\begin{eqnarray*}
\mbox{Br}_{LC}\left(B_{c}\to J/\psi\pi\right) & = & 0.13\%,\\
\mbox{Br}_{QM}\left(B_{c}\to J/\psi\pi\right) & = & 0.17\%,\\
\mbox{Br}_{SR}\left(B_{c}\to J/\psi\pi\right) & = & 0.17\%.\end{eqnarray*}

\subsection{$B_{c}\to J/\psi+2\pi$}

The $2\pi$ channel is saturated mainly by $B_{c}\to J/\psi\rho$
decay. The $W^{*}\to\rho$ transition vertex is also expressed through
$\rho$-meson leptonic constant

\begin{eqnarray*}
\left\langle \rho\left|\bar{u}\gamma_{\mu}d\right|0\right\rangle  & = & \sqrt{2}f_{\rho}M_{\rho}\epsilon_{\mu}\end{eqnarray*}
where $f_{\rho}\approx150\,\mbox{MeV}.$ If one neglects the width
of $\rho$-meson, the corresponding spectral function has the form

\begin{eqnarray}
\rho_{T}^{\rho}\left(q^{2}\right) & = & 2f_{\rho}^{2}\delta\left(q^{2}-m_{\rho}^{2}\right).\label{eq:rhoDelta}\end{eqnarray}
The branching fractions of $B_{c}\to J/\psi\rho$ for different sets
of form-factors are:\begin{eqnarray*}
\mbox{Br}_{LC}\left(B_{c}\to J/\psi\rho\right) & = & 0.38\%,\\
\mbox{Br}_{QM}\left(B_{c}\to J/\psi\rho\right) & = & 0.44\%,\\
\mbox{Br}_{SR}\left(B_{c}\to J/\psi\rho\right) & = & 0.48\%.\end{eqnarray*}

\begin{figure}
\begin{centering}
\includegraphics[scale=0.7]{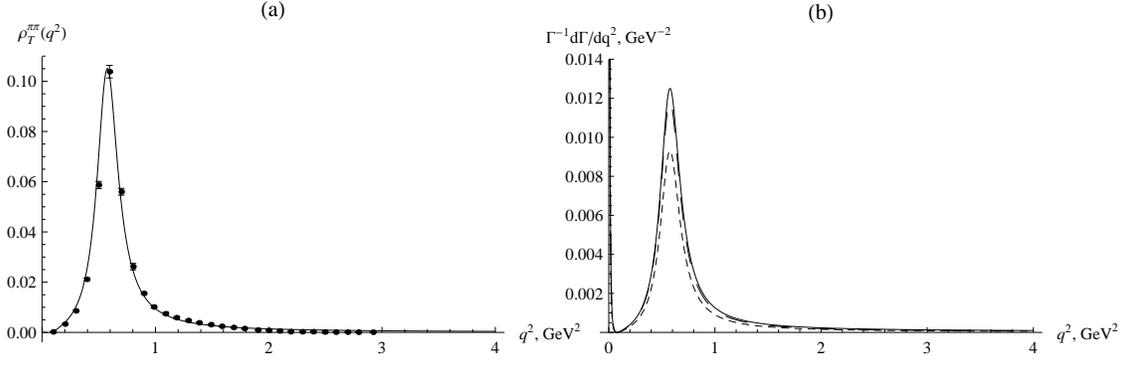} 
\par\end{centering}

\caption{fig (a) --- spectral function $\rho_{T}^{2\pi}$; (b) --- $\Gamma^{-1}d\Gamma(B_{c}\to J/\psi+2\pi)/dq^{2}$
distribution for different sets of $B_{c}$-meson form-factors. Solid,
dashed and dot-dashed lines stand for SR \cite{Kiselev2000pp,Kiselev1999sc},
QM \cite{Gershtein1995}, and LC \cite{Wang2009mi} respectively\label{fig:2pi}}

\end{figure}

In order to take $\rho$-meson width into account, one can use experimental
data on $\tau\to\nu_{\tau}+2\pi$ decay. The differential branching
ratio of this reaction is equal to\begin{eqnarray*}
\frac{d\Gamma\left(\tau\to\nu_{\tau}\mathcal{\mathcal{R}}\right)}{dq^{2}} & = & \frac{G_{F}^{2}}{16\pi m_{\tau}}\frac{\left(m_{\tau}^{2}-q^{2}\right)^{2}}{m_{\tau}^{3}}\left(m_{\tau}^{2}+2q^{2}\right)\rho_{T}^{\mathcal{R}}\left(q^{2}\right).\end{eqnarray*}
This method was used by ALEPH collaboration to measure the spectral
function $\rho_{T}^{2\pi}(q^{2})$ in the kinematically allowed region
$q^{2}<m_{\tau}^{2}$ \cite{Schael2005} and can be approximated by
the expression (see fig.\ref{fig:2pi}a)\begin{eqnarray*}
\rho_{T}^{2\pi}\left(s\right) & \approx & 1.35\times10^{-3}\left(\frac{s-4m_{\pi}^{2}}{s}\right)^{2}\frac{1+0.64s}{\left(s-0.57\right)^{2}+0.013},\end{eqnarray*}
where $s$ is measured in $\mbox{GeV}^{2}$. In fig.\ref{fig:2pi}b
we show corresponding distributions $d\Gamma\left(B_{c}\to J/\psi+2\pi\right)/dq^{2}$.
Solid, dashed and dash-dotted lines in this figure correspond to form-factors
SR, QM, and LC respectively. The branching fractions of the decay
$B_{c}\to J/\psi+2\pi$ are almost equal to $B_{c}\to J/\psi\rho$
decay branching fractions:\begin{eqnarray*}
\mbox{Br}_{LC}\left(B_{c}\to J/\psi\pi\pi\right) & = & 0.35\%,\\
\mbox{Br}_{QM}\left(B_{c}\to J/\psi\pi\pi\right) & = & 0.44\%,\\
\mbox{Br}_{SR}\left(B_{c}\to J/\psi\pi\pi\right) & = & 0.48\%.\end{eqnarray*}

\subsection{$B_{c}\to J/\psi+3\pi$}

\begin{figure}
\begin{centering}
\includegraphics[scale=0.7]{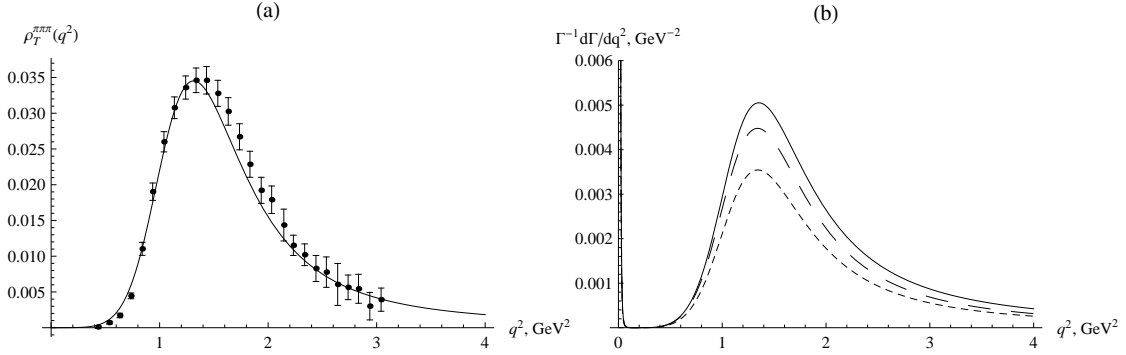} 
\par\end{centering}

\caption{Spectral function and differential width for $B_{c}\to J/\psi+3\pi$
decay. Notations are the same as in fig.\ref{fig:2pi}.\label{fig:3pi}}

\end{figure}

In the case of $B_{c}\to J/\psi+3\pi$ decay (where $3\pi$ stands
for the sum of $\pi^{-}\pi^{0}\pi^{0}$ and $\pi^{-}\pi^{+}\pi^{-}$
decay modes) the $G$-parity of the final state is negative. So we
can expect, that this mode is saturated by axial-vector resonance
$a_{1}$. The width of this state is too large to neglect it, so we
cannot use the expression similar to (\ref{eq:rhoDelta}) for $W^{*}\to3\pi$
transition. The corresponding spectral function can be determined
from experimental and theoretical data on $\tau\to\nu_{\tau}+3\pi$
decay. In our article we use the following expression to approximate
this function (see. fig.\ref{fig:3pi}a):\begin{eqnarray*}
\rho_{T}^{3\pi}\left(s\right) & \approx & 5.86\times10^{-5}\left(\frac{s-9m_{\pi}^{2}}{s}\right)^{4}\frac{1+190.s}{\left[\left(s-1.06\right)^{2}+0.48\right]^{2}}.\end{eqnarray*}
Distributions over $q^{2}$ for different sets of $B_{c}$-meson form
factors are shown in fig.\ref{fig:3pi}b. The branching fractions
of $B_{c}\to J/\psi+3\pi$ decay are\begin{eqnarray*}
\mbox{Br}_{LC}\left(B_{c}\to J/\psi+3\pi\right) & = & 0.52\%,\\
\mbox{Br}_{QM}\left(B_{c}\to J/\psi+3\pi\right) & = & 0.64\%,\\
\mbox{Br}_{SR}\left(B_{c}\to J/\psi+3\pi\right) & = & 0.77\%.\end{eqnarray*}

\subsection{$B_{c}\to J/\psi+4\pi$}

\begin{figure}
\begin{centering}
\includegraphics[scale=0.7]{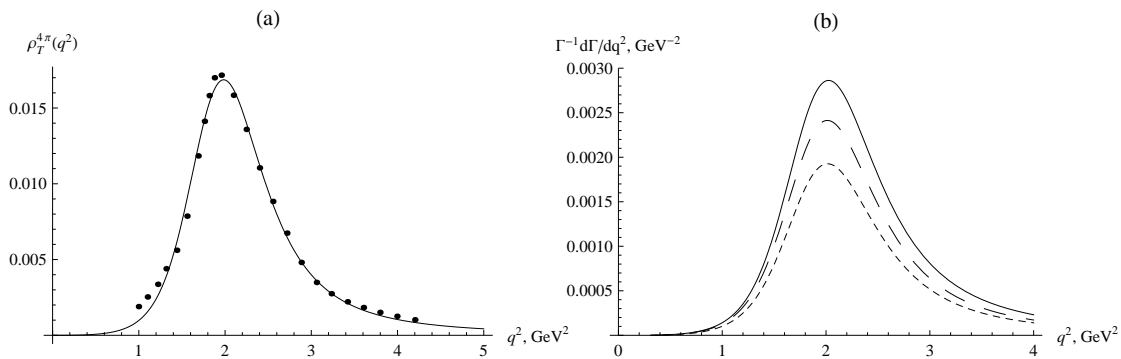} 
\par\end{centering}

\caption{Spectral function and differential width for $B_{c}\to J/\psi+4\pi$
decay. Notations are the same as in fig.\ref{fig:2pi}\label{fig:4pi}}

\end{figure}

In the decay $B_{c}\to J/\psi+4\pi$ both $\pi^{-}\pi^{0}\pi^{0}\pi^{0}$
and $\pi^{-}\pi^{+}\pi^{-}\pi^{-}$ modes are possible in the following
we consider the sum of these states. The kinematically allowed region
in $\tau\to\nu_{\tau}+4\pi$ decay is too small to determine the form
of spectral function $\rho_{T}^{4\pi}$, so it is more convenient
to use energy dependence of $4\pi$ production cross section in electron-positron
annihilation. It is easy to obtain the following expression for this
cross section:\begin{eqnarray*}
\sigma\left(e^{+}e^{-}\to4\pi\right) & = & \frac{4\pi\alpha^{2}}{s}\rho_{T}^{4\pi}(s).\end{eqnarray*}
Spectral function $\rho_{T}^{4\pi}$, calculated from experimental
data \cite{Czyz2001} is shown in fig.\ref{fig:4pi}a and later we
use the following parametrization:\begin{eqnarray*}
\rho_{T}^{4\pi}\left(s\right) & \approx & 1.8\times10^{-4}\left(\frac{s-16m_{\pi}^{2}}{2}\right)\frac{1+5.07s+8.63s^{2}}{\left[\left(s-1.83\right)^{2}+0.61\right]^{2}}.\end{eqnarray*}
The distributions corresponding to this spectral function are shown
in fig.\ref{fig:4pi}b. The branching fraction for different sets
of $B_{c}$-meson form-factors are\begin{eqnarray*}
\mbox{Br}_{LC}\left(B_{c}\to J/\psi+4\pi\right) & = & 0.26\%,\\
\mbox{Br}_{QM}\left(B_{c}\to J/\psi+4\pi\right) & = & 0.33\%,\\
\mbox{Br}_{SR}\left(B_{c}\to J/\psi+4\pi\right) & = & 0.40\%.\end{eqnarray*}

\section{Inclusive Decays and Duality Relation}

\begin{figure}
\begin{centering}
\includegraphics{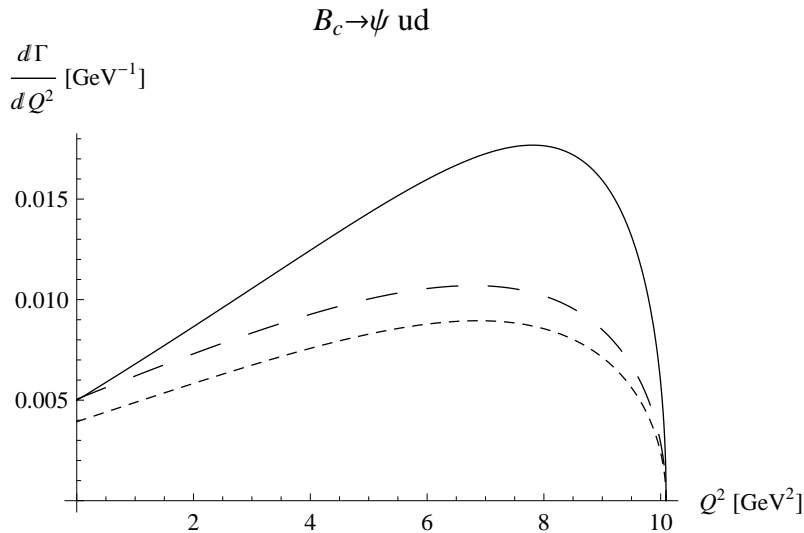}
\par\end{centering}

\caption{Differential $B_{c}\to J/\psi\bar{u}d$ branching fractions for different
sets of $B_{c}$-meson form-factors. Notations are the same as in
fig.\ref{fig:2pi}\label{fig:ud}}

\end{figure}

Let us know consider the inclusive decay $B_{c}\to B_{c}+X$ where
$X$ stands for an arbitrary state of light hadrons. On quark level
this reaction corresponds to $B_{c}\to J/\psi+\bar{u}d$ decay. If
one neglects $u$- and $d$-quark masses, the spectral function of
$W^{*}\to\bar{u}d$ transition is energy independent and equals to\begin{eqnarray*}
\rho_{T}^{ud} & = & \frac{1}{2\pi^{2}}.\end{eqnarray*}
In fig.\ref{fig:ud} distributions of $B_{c}\to J/\psi+\bar{u}d$
decay branching fractions for different sets of $B_{c}$-meson form-factors
are shown. Integrated branching fractions of this decay are\begin{eqnarray*}
\mbox{Br}_{LC}\left(B_{c}\to J/\psi+\bar{u}d\right) & = & 7\%,\\
\mbox{Br}_{QM}\left(B_{c}\to J/\psi+\bar{u}d\right) & = & 8.6\%,\\
\mbox{Br}_{SR}\left(B_{c}\to J/\psi+\bar{u}d\right) & = & 12\%.\end{eqnarray*}
It should be noted, that sum of presented above branching fractions
(that is $Br(B_{c}\to J/\psi+n\pi)$, $n=1,\dots4$) gives only about
30\% the inclusive decay branching fraction. So one could expect noticeable
events with multi-pion production in $B_{c}$-meson decays.

Below $KK$-production threshold only $\pi$-mesons can be produced
in $\bar{u}d$-pair hadronization, so the duality relation should
be satisfied\begin{eqnarray*}
\int\limits ^{(2m_{K}+\Delta)^{2}}\frac{1}{\Gamma}\frac{d\Gamma(B_{c}\to J/\psi\bar{u}d)}{dq^{2}} & = & \sum_{n}\mbox{Br}(B_{c}\to J/\psi+n\pi),\end{eqnarray*}
where $\Delta$ is the duality window. If we restrict ourselves to
$n\le4$ in the right-hand side of this relation, it is valid for\begin{eqnarray*}
\Delta & \approx & 0.6\,\text{GeV}.\end{eqnarray*}
It is interesting to note that this value is almost independent on
the choice of $B_{c}$-meson form-factors and close to the value of
duality parameter in $gg\to J/\psi c\bar{c}$ and $\chi_{b}\to J/\psi c\bar{c}$
reactions \cite{Berezhnoy2006mz,Braguta2005pp}.

\section{Conclusion}

In our paper we study exclusive and inclusive decays of $B_{c}$-meson
into light hadrons and vector charmonium $J/\psi$, that is the processes
$B_{c}\to J/\psi+\bar{u}d$ and $B_{c}\to J/\psi+n\pi$ where $n=1,2,3,4$.
According to QCD factorization theorem the amplitude of these processes
splits into two independent parts. The first factor describes the
decay $B_{c}\to J/\psi W^{*}$ and one can use existing parametrizations
of $B_{c}$-meson form-factors to calculate this amplitude. The second
factor describes the fragmentation of virtual $W$-boson. The information
about these processes was taken from experimental distributions of
multi-pion production in $\tau$-lepton decays and electron-positron
annihilation.

Our results are gathered in table \ref{tab:Br}, where branching fractions
of multi-pion production in $B_{c}\to J/\psi+n\pi$ for different
$B_{c}$-meson form-factors are presented. The last column of this
table contains the branching fraction of the inclusive decay $B_{c}\to J/\psi+\bar{u}d$.
It is clear that up to $KK$-production threshold only $\pi$-mesons
could be produced in $B_{c}\to J/\psi+X$ decay, so some duality relation
should hold. In our article it is shown, that to satisfy this relation
it is sufficient to integrate the inclusive spectrum up to squared
transferred momentum $q^{2}=(2m_{K}+\Delta)^{2}$. It turns out, that
$\Delta$ is almost independent on the choice of $B_{c}$-meson form-factors
and equals to $\sim0.6$ GeV.

\begin{table}
\begin{centering}
\begin{tabular}{|c|c|c|c|c|c|}
\hline 
 & $\pi$ & $2\pi$ & $3\pi$ & $4\pi$ & $\bar{u}d$\tabularnewline
\hline
\hline 
$LC$ & 0.13 & 0.35 & 0.52 & 0.26 & 7\tabularnewline
\hline 
$QM$ & 0.17 & 0.44 & 0.64 & 0.33 & 8.6\tabularnewline
\hline 
$SR$ & 0.17 & 0.48 & 0.77 & 0.40 & 12\tabularnewline
\hline
\end{tabular}
\par\end{centering}

\caption{$B_{c}\to J/\psi\mathcal{R}$ decays branching fractions (in \%) for
different sets of $B_{c}$-meson form-factors\label{tab:Br}}

\end{table}

The other interesting point are the polarization asymmetries of final
$J/\psi$-meson. In the framework of factorization model these asymmetries
do not depend on the final state $\mathcal{R}$, so one can use them
to investigate form-factors of $B_{c}$-meson and to test the factorization
theorem. In our paper we present the polarization degree $\alpha=(d\Gamma_{T}/dq^{2}-2d\Gamma_{L}/dq^{2})/(d\Gamma_{t}/dq^{2}+2d\Gamma_{L}/dq^{2})$
and transverse polarization asymmetry $\alpha_{T}=(d\Gamma_{\lambda=1}/dq^{2}-d\Gamma_{\lambda=-1}/dq^{2})/(d\Gamma/dq^{2})$
for different sets of form-factors.

The authors would like to thank V.V. Kiselev for fruitful discussions.
This work was financially supported by Russian Foundation for Basic
Research (grants \#09-02-00132-a and 07-02-00417-a). One of the author
(A.V.L.) was also supported by President grant (\#MK-110.2008.2),
grant of Russian Science Support Foundation and noncommercial foundation
\textquotedblright{}Dynasty\textquotedblright{}.

\end{document}